# Variations of two-neutron separation energies and thermal-neutron capture cross-sections versus the pairing gap

Hossein Emami[1*], Hadi Sabri[1]

[1]Faculty of Physics, University of Tabriz, Tabriz, Iran

Corresponding author e-mail: husseinnuclear92@gmail.com

Abstract

In this work, we investigate the experimental correlation between the pairing gap values and two important observables in the study of nuclear structure (two neutron separation energies and thermal-neutron capture cross-sections). To this aim, we focused on the even-even nuclei in the vicinity of Z=50 and Z = 82 closed proton shells, for which the quantum phase transition phenomena are reported. The results show a significant correlation between the pairing gap and the well-known signatures of quantum phase transitions in the nuclei, which are the candidates for E(5) and X(5) critical points. Also, we have explained the special relation between the pairing gap and the cross-section of thermal neutrons in the considered isotopic chains.

1. Introduction

Quantum phase transition (QPT) refers to the changes in the fundamental structure of atomic nuclei at absolute zero. The phenomenon known as a shape phase transition(SPT), also referred to as a ground state phase transition, can also be observed in excited states in addition to its occurrence in the ground state [1-4]. The discontinuity occurs in the first-order case for the first energy derivative, while it applies to the second-order derivative in the second-order case. The first-order SPT indicates an abrupt change compared to the second-order transitions [5] .There are some investigations about signatures of QPT or parameters that can be investigated in the search for QPT, such as two neutron separation energy ($S_{2n}$) [5-13], β-γ order parameters [14-17], thermal-neutron capture cross-section ($\sigma_{th}$) [18], and among others. The observables mentioned above are macroscopic signatures for QPT, and we aim to relate macroscopic signatures to the microscopic concepts. Shape coexistence refers to the specific situation in which the ground state band of the nucleus is close to another band with a completely different structure. In even-even nuclei, shape coexistence often leads to the presence of a $0^+$ band that is closely situated in energy to the ground state band yet possesses a fundamentally different structure. For instance, one of the bands may be spherical while the other is deformed. Thus, the nature of low-lying $0^+$ bands in even-even nuclei is of interest [19]. Most studies have focused on experimental observations and relevant theoretical developments [19-21]. QPT and Shape coexistence can be related to each other, and some investigations about the connection between the QPT and Shape coexistence have been given in

Ref. [5,19,22-24]. The energy gap between the ground state and the nearly degenerate states is approximately 1 MeV. The pairing gap plays a crucial role in the characteristics of nuclear structures. Atomic nuclei can undergo deformation, which is influenced by various factors such as nuclear structure features, compound effects, pairing gap effects, and quadrupole-quadrupole (Q-Q) interaction. The pairing and Q-Q interactions are the most important short-range and long-range correlations, respectively [25]. When the pairing gap becomes larger, the nucleus tends to become more spherical. Conversely, increasing the Q-Q interaction results in a deformed state. As a result, the shape of the nucleus is influenced in favor of whichever parameter, the pairing gap or the Q-Q interaction, is more dominant. When we deviate from the magic numbers and head towards the critical points, the pairing gap shifts from its stable state to a deformed state and then transitions to the next symmetrical state. The study of transitions from one phase to another was provided by this fact, which led to the creation of critical point symmetry for these phase transitions [26]. In nuclear physics, there are two critical point symmetries known as the E(5) [27] and X(5) symmetries [28]. In some studies, the pairing gap has been partially used to describe the SPT in the nucleus [29]. Recently, various studies have been conducted on the effects of the pairing gap on different phenomena [29-33]. Also, we will show in the following that the pairing gap can be a parameter similar to QPT that has some observables such as $S_{2n}$ and $\sigma_{th}$.

The excitation spectrum is defined by the existence of an energy gap [34]. In atomic nuclei, the pairing gap is generally about 1-2 MeV, which is much smaller than the typical energy scale of the N-N interaction, which is a few hundred MeV [35]. The pairing gaps between nucleons were determined using formulas based on the even-odd mass differences, which are part of the liquid-drop term explained by Bohr and Mottelson [36]:

$$\Delta_{BM} \approx 12 A^{-\frac{1}{2}}, \tag{1}$$

Also, the pairing gaps are determined using the binding energy and separation energies according to the formulas:

$$\Delta(Z,N) = \frac{(-1)^N}{2}[2BE(Z,N) - BE(Z,N-1) - BE(Z,N+1)], \tag{2}$$

$$\Delta(Z,N) = \frac{(-1)^{N+1}}{2}[S_n(Z,N+1) - S_n(Z,N)], \tag{3}$$

The effective pairing interaction at the Fermi surface is challenging to control due to its microscopic construction. The Bardeen, Cooper, and Schrieffer (BCS) [37] approach often depends on an approximation known as the weak coupling limit. The value of the pairing gap is strongly affected by the density of states at the Fermi level, as indicated by the weak coupling limit [35]. Commonly, the pairing gap in a nucleus is often estimated using its spectral properties. However, specific definitions of this pairing gap may not apply to closed-shell nuclei [34]. In the near magic number, the pairing gap increases to its maximum value. This value indicates the important concepts of nuclear structures, including strong binding between two fermions, more stability against reactions, long half-life, spherical state, and others. Thus, the pairing gap is related to the above concepts directly, which means the pairing gap is associated with $S_{2n}$, $\sigma_{th}$, and QPT at the same time.

Nuclear masses and binding energies, more specifically $S_{2n}$, are crucial observables characterizing a nucleus and providing insights into nuclear correlations. $S_{2n}$, within the IBM framework, is a valuable tool for explaining the observed phase transitions in even-even mass nuclei. This quantity was proposed in a paper that studied the classical limit of IBM [38], and it has proven to be one of the most useful measures in this context. $S_{2n}$ and their evolution with neutron number provide an excellent basis for testing various nuclear structure models [10]. Therefore, $S_{2n}$ can be one of the most robust observables of QPT [5-13].

Two main types of neutron capture reactions are slow neutron capture (s-process) and rapid neutron capture (r-process). The s-process focuses on stable and near-stable nuclei, competing with β decay, while the r-process progresses beyond β decay towards neutron-rich nuclei until specific conditions apply. A nucleus with weak coupling and binding energy is more exposed to reactions, and that means this nucleus has a large cross-section to capture thermal-neutron in the external interaction, and this leads to instability, short half-life, deformation states, and others. A recent discovery involves a purely empirical correlation between $S_{2n}$ values and neutron capture cross-sections at neutron energies relevant to nucleosynthesis [39,40]. Also, it has been shown in Ref. [18] that $\sigma_{th}$ can be one of the new observables of QPT. Two observables ($\sigma_{th}$, $S_{2n}$) due to the distribution of nuclear matter and nuclear structure in general. These observables correlated to neutron pairing; thus, according to this, we investigated these observables versus the pairing gap. We start our investigation about a possible connection between the pairing gap and ($\sigma_{th}$, $S_{2n}$) first, so we will conclude that the pairing gap can be related to these observables as a more sensitive parameter in the study of nuclear structures and extract some features about this investigation as applications of the pairing gap finally.

## 2. Results and discussion

Our investigations in this paper are based on two sections in general: 1. correlation between the pairing gap and $S_{2n}$, and 2. correlation between the pairing gap and $\sigma_{th}$. We also will use just purely experimental data about the pairing gap, $S_{2n}$, and $\sigma_{th}$ because the final aim of this article is to investigate the possible relation between the pairing gap and ($S_{2n}$, $\sigma_{th}$). $S_{2n}$ as an observable of QPT has been used in most studies, while $\sigma_{th}$ is a new observable that has been investigated about QPT in Ref. [18] recently.

### 2.1. Section 1: Correlation between the pairing gap and $S_{2n}$

$S_{2n}$ is an observable that represents both types of QPT (first/second-order). The energy required to remove two neutrons from a nucleus is also known as:

$$S_{2n} = 2M_n + M(Z, N-2) - M(Z, N), \qquad (4)$$

Where M(N, Z) is the mass of nuclei with neutrons and protons, respectively, and $M_n$ is the neutron mass. Also, the separation energies for two neutrons are calculated by the difference of binding energy and one neutron separation energy of two isotopes using relations:

$$S_{2n}(Z, N) = BE(Z, N) - BE(Z, N-2), \qquad (5)$$

$$S_{2n}(Z, N) = S_n(Z, N-1) + S_n(Z, N), \qquad (6)$$

Therefore, we investigate the correlation between the pairing gap and $S_{2n}$ first, and according to that, $S_{2n}$ is one of the important macroscopic signatures for QPT. We want to show that this observable is due to the pairing gap. According to the relation between one/two neuron separation energies and binding energies, and also according to the relation between the pairing gap and binding energies, we can define a new relation between the pairing gap and one/two neuron separation energies, which is given by the following:

$$\Delta(Z,N) = \frac{(-1)^{N+1}}{2}[S_{2n}(Z,N+1) - 2S_{2n}(Z,N) + 2S_n(Z,N-1)], \qquad (7)$$

Eq. (7) can be directly derived from Eq. (6), and it does not provide new information to better understand the pairing gap, but our aim in providing this formula is to show that the pairing gap and $S_{2n}$ can be used in general form. Our focus in this section is on the evolution of the pairing gap as a function of $S_{2n}$, by using empirical data (taken from [41]). We will use even-even isotopic chains in the near shell closure Z=50 and Z=82 respectively, these isotopic chains are: Ru, Pd, Cd, Sn, Os, Pt, Hg, and Pb. The evolution of the pairing gap versus $S_{2n}$ is shown in Fig. 1 and the evolution of $S_{2n}$ versus neutron number, in terms of the pairing gap by contour-color filling method, is shown in Fig. 2. It should be noted that we used just the experimental data (taken from [42]), about the pairing gap and $S_{2n}$.

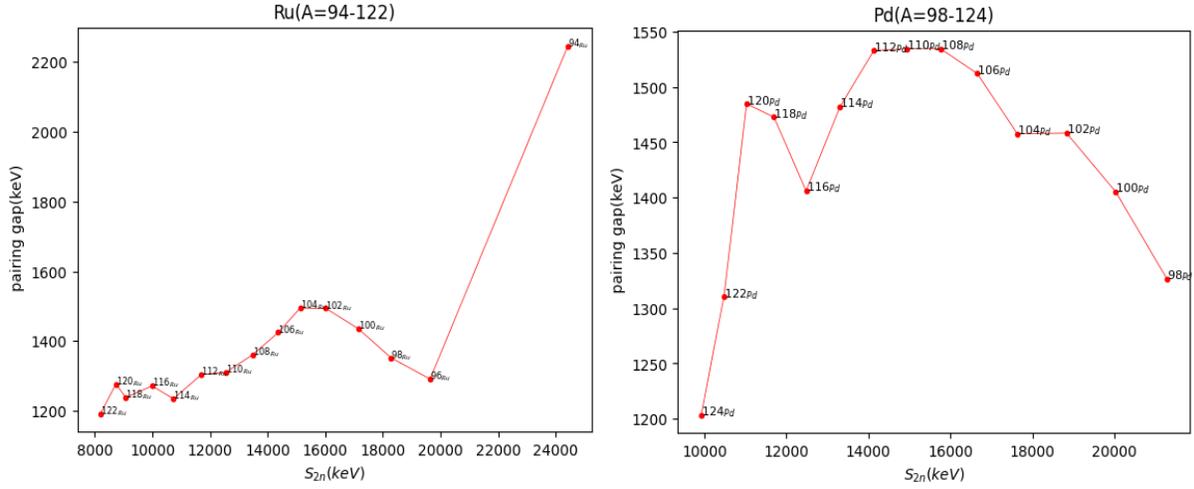

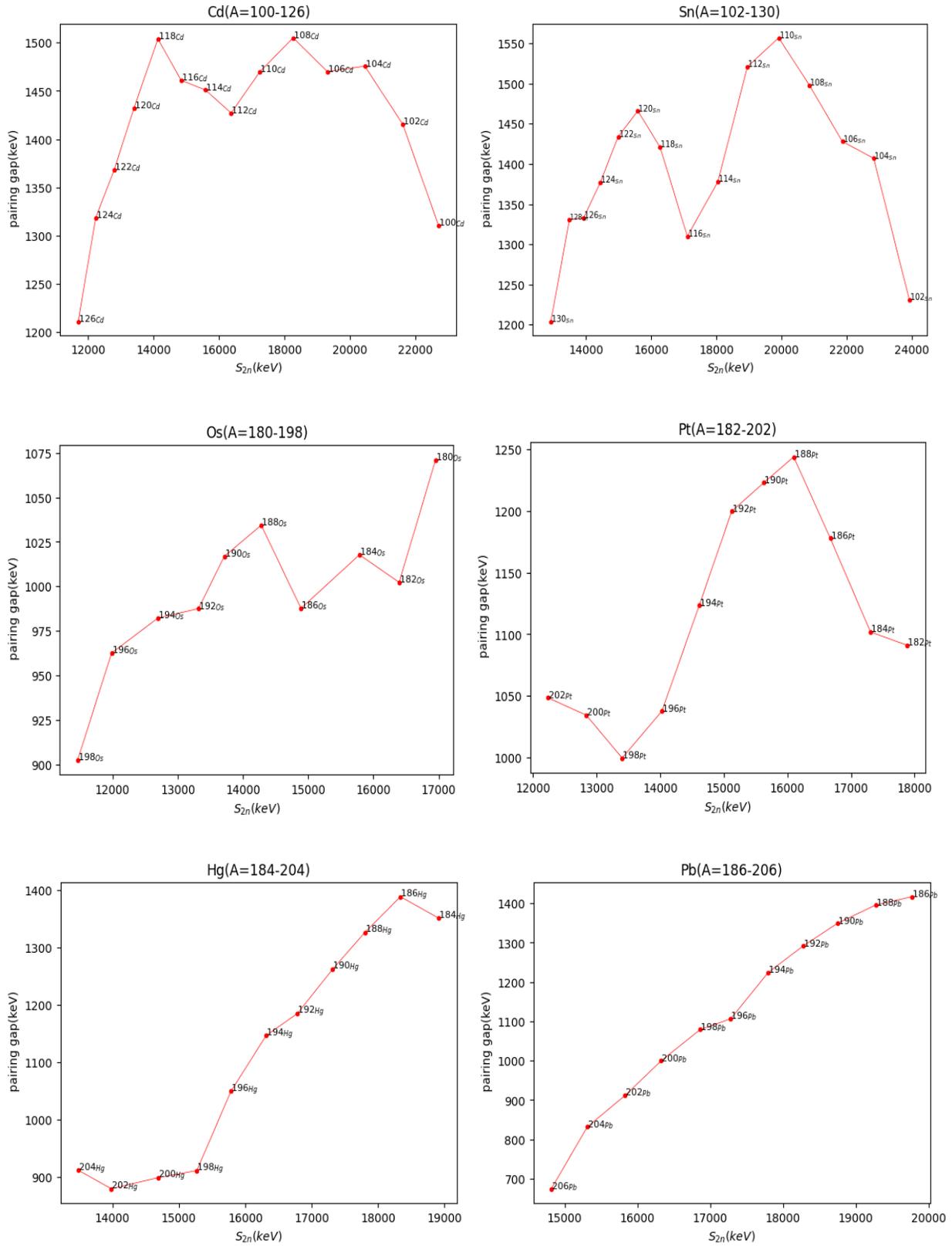

Figure 1. The variation of the pairing gap versus $S_{2n}$, for Ru, Pd, Cd, Sn, Os, Pt, Hg, and Pb isotopic chains.

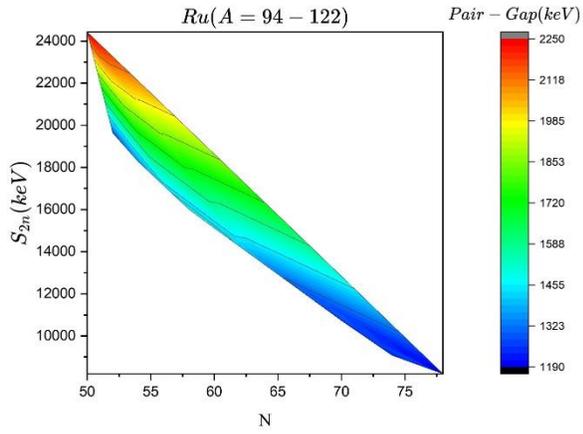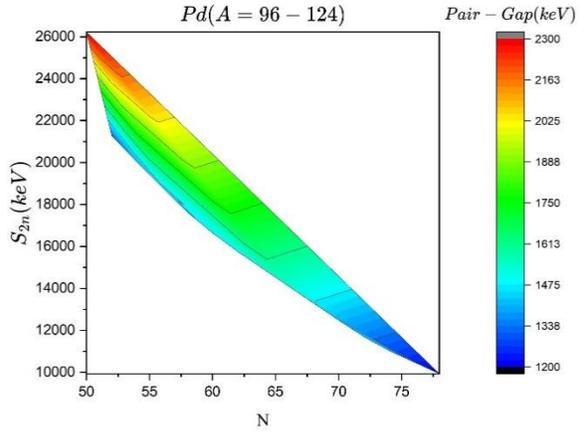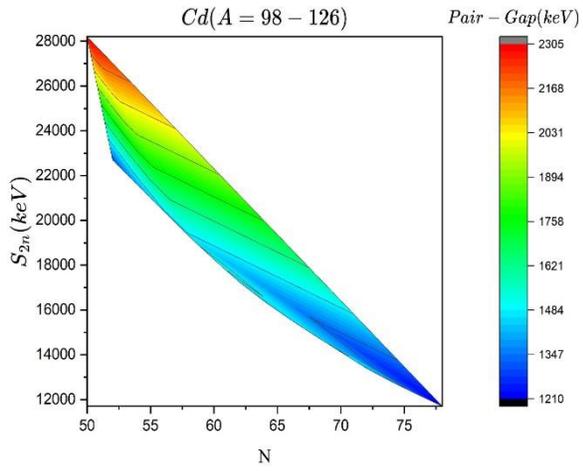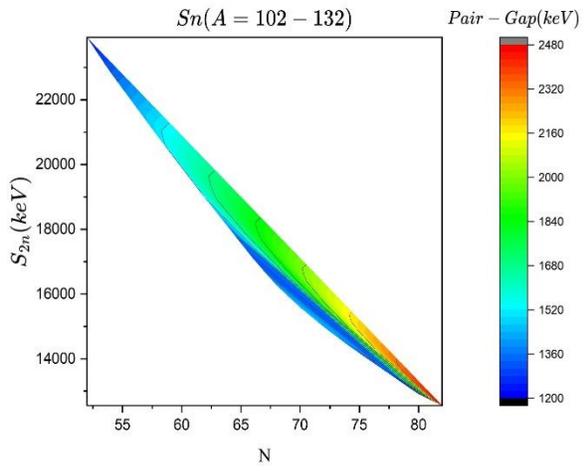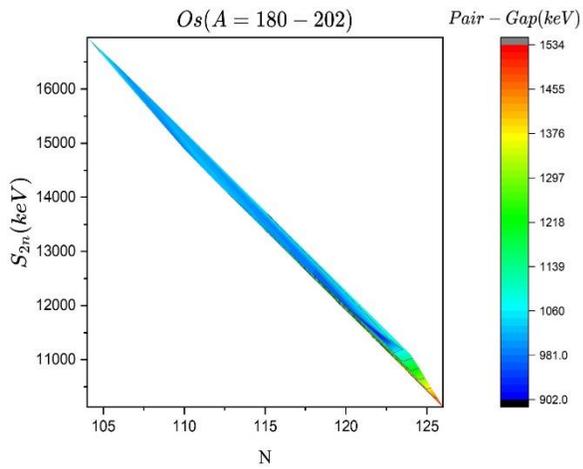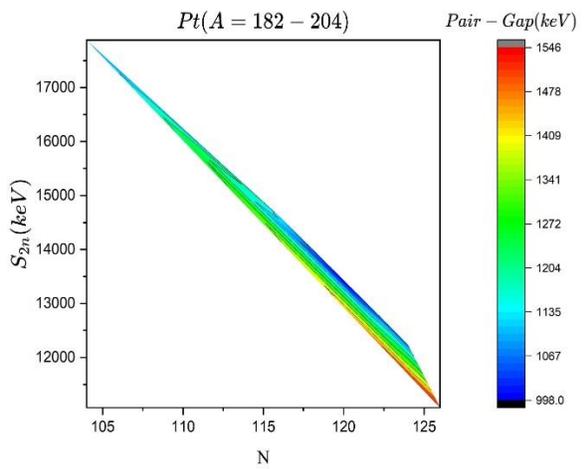

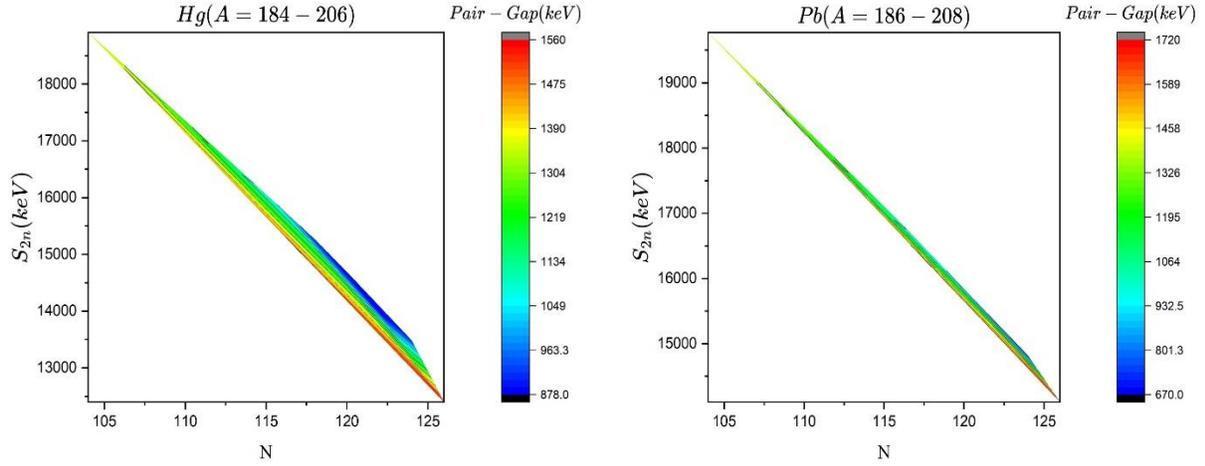

Figure 2. Contour plots in the ($S_{2n}$, neutron number) plane, in terms of the pairing gap, for Ru, Pd, Cd, Sn, Os, Pt, Hg, and Pb isotopic chains.

We used the contour plots method in the ($S_{2n}$, neutron number) plane, in terms of the pairing gap to show the inequality of these quantities, on the other hand, we want to identify that the pairing gap comes from the difference between two $S_{2n}$, that's between an isotope and its adjacent isotope. We didn't put magic numbers in (Fig. 1), except for the Ru isotopic chain, because in magic numbers, the pairing gap value takes a large peak and makes it difficult for us to analyze the graphs. According to Fig. 1 and Fig. 2, the pairing gap's energy level in the near shell closure Z=50, is higher than near shell closure Z=82, at the same $S_{2n}$ range, because, in the near shell closure Z=50, we are near to drip-line, thus not surprising that, stability in the near shell closure Z=50 is more than near shell closure Z=82. In the near shell closure Z=50, when $S_{2n} \approx 15500 - 16000 \ keV$, the pairing gap values begin to decrease, and similar Ref. [10,13], it's possible, that isotopes within this $S_{2n}$ range are critical points symmetry in nuclei, including ($^{104,102}Ru$ [43,44], $^{108,110}Pd$ [45,46], $^{114}Cd$ [47]), and according to these have been reported candidates for E(5) and X(5) critical points, we suggest $^{120}Sn$ can be as a new critical point for Sn isotopic chains and, we have no claim that our candidate is a critical point, we just suggested a candidate based on a special range of $S_{2n}$, also this nuclei has $S_{2n}$ in range of $\approx 15500 - 16000 \ keV$. This range represents nearly double the average binding energy. The final aim in graphs of isotopic chains that are near shell closure Z=50 (Fig. 1) is to reach the double peaks state(M-shaped mode), so that $\frac{(S_{2n})_{peak_2}}{(S_{2n})_{peak_1}} \approx 1.28$ and isotopes which are located in these peaks, have a special ratio: $\frac{(R_{4_1^+/2_1^+})peak_2}{(R_{4_1^+/2_1^+})peak_1} \approx 1$, such as ($^{110,120}Pd$, $^{108,118}Cd$, $^{110,120}Sn$), therefore, the enhanced pairing gaps in these nuclei are related to some shell effects and also have been observed shape coexistence in the different study for these nuclei in Ref. [19,48]. In this order, these peaks are related to different shells in the nuclear structure, on the other hand, the first peak indicates ($^{120}Pd$, $^{118}Cd$, $^{120}Sn$), which are located after mid-shell (N=66) also, these nuclei have a low pairing gap values than which nuclei are located in second peak that including ($^{110}Pd$, $^{108}Cd$, $^{110}Sn$), and these nuclei are located in before mid-shell (N=66). Thus, these double peaks state in near shell closure Z=50 due to the crossing from mid-shell to others, and we will investigate some classification of nuclei versus the pairing gaps based on some shell effects in future works. The ups and downs of graphs are due to the secondary effects such as Q-Q interaction, spin, etc, according to Fig. 1, in the near shell closure Z=82, the graphs of isotopic chains reach the linear state finally because the secondary effects decrease with the increasing atomic number in the near shell closure Z=82, therefore just the pairing gap remains to show the evolution of $S_{2n}$ and this means that $S_{2n}$ as the observable of QPT can be due to the pairing gap. According to Fig. 2, when we move towards the magic number N=50, both the pairing gap and $S_{2n}$, increase, while by moving towards the magic numbers N=82 and N=126, just the pairing gap increases and $S_{2n}$ decreases. By increasing the atomic number, the pairing gap's energy levels increase in each of the shell closures because the secondary effects decrease by increasing the atomic number.

## 2.2 Section 2: Correlation between the pairing gap and $\sigma_{th}$

In this section, we investigate $\sigma_{th}$ ($kT \simeq 30\ keV$) and its correlation with the pairing gap value by using purely empirical data (taken from [41, 42]).

Some experimental relations for $\sigma_{th}$ recently discovered, like its relation with neutron energies and $S_{2n}$ [39,40], as:

$$\sigma_{th} = \sigma_{E_n} C_0 (E_n)^{C_1}, \tag{8}$$

Where $E_n$ is the Maxwell-Boltzmann average energy in keV, $C_0$ and $C_1$ were fit to the rare-earth even-even data to give $C_0 = 0.154$ and $C_1 = 0.552$.

$$\sigma_{fit} = p_0 \cdot e^{p_1 \cdot (p_2 + S_{2n})}, \tag{9}$$

where $p_0$ is 0.0106±0.0005, $p_1$ is 0.7644±0.0035, and $p_2$ is zero for even-even deformed nuclei in the rare earth region. For transitional even-even nuclei in the rare earth region $p_2 = -1.92$ [39].

According to sect. 1, we also use even-even isotopic chains in the near shell closure Z=50 and Z=82, respectively, in this section, and these isotopic chains are Ru, Pd, Cd, Sn, Os, Pt, Hg, and Pb. In Addition to these isotopic chains, we also examine Sm and Gd isotopic chains in rare earth around A=150 as application of the pairing gap, by contour-color filling method, shown in Fig. 5

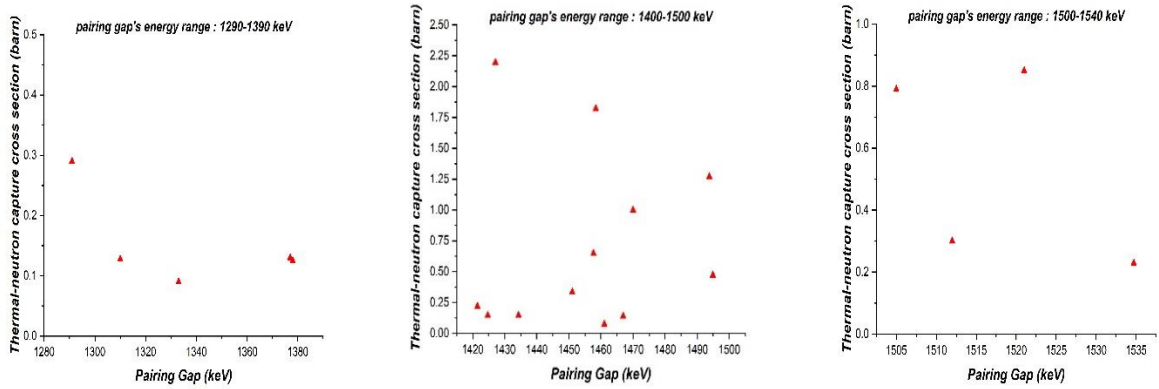

Figure 3. The variation of $\sigma_{th}$ versus the pairing gap in different ranges of the pairing gap's energy for even-even isotopic chains that near shell closure Z=50 (Ru, Pd, Cd, Sn)

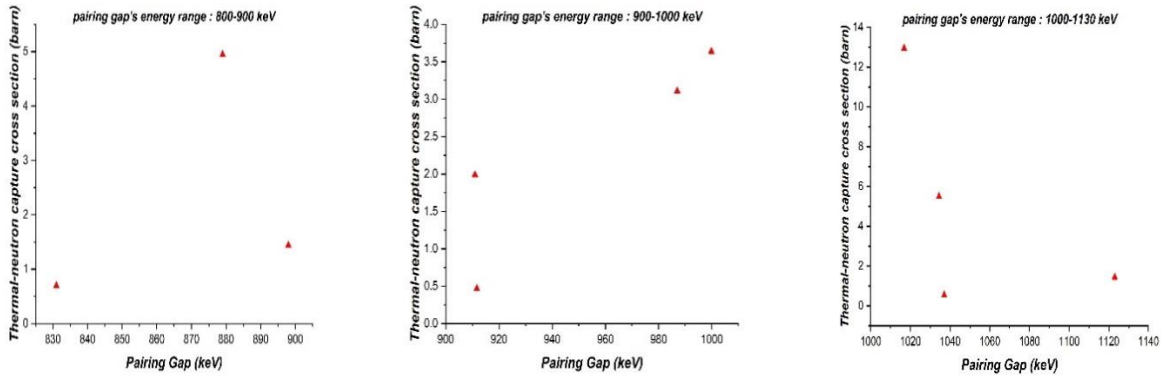

Figure 4. The variation of $\sigma_{th}$ versus the pairing gap in different ranges of the pairing gap's energy for even-even isotopic chains in the near shell closure Z=82 (Os, Pt, Hg, Pb)

The first purpose of this section is to investigate different cross-sections that are located in the specific range of the pairing gap value, shown in Fig. 3 and Fig. 4. In other words, which thermal neutrons are in the specific 'gap' and will capture in the nucleus? According to Fig. 3, in the near shell closure Z=50, most isotopes tend to capture the thermal neutrons with cross-sections 0.07-2.25 (barn), in the pairing gap 1400-1500 (keV) range. Knowing that, the pairing gap's energy level in the near shell closure Z=50, is higher than near shell closure Z=82, and according to Fig. 3 and Fig. 4, neutron capture cross-sections in the near shell closure Z=82, are higher than near shell closure Z=50, that's the isotopes tend to capture the thermal-neutrons with higher cross-sections in the near shell closure Z=82 than near shell closure Z=50, by a different range of the pairing gap's energy. We aim to conclude and analyze data from selected isotopic chains globally. Because the pairing gap and $\sigma_{th}$ are due to the two different interactions in atomic nuclei, the pairing gap is related to the internal interaction in nuclei, and $\sigma_{th}$ is related to external interaction along gamma-ray production. Therefore, by this global investigation between shell closures, the inverse relation between the pairing gap and $\sigma_{th}$ in these isotopic chains can be seen.

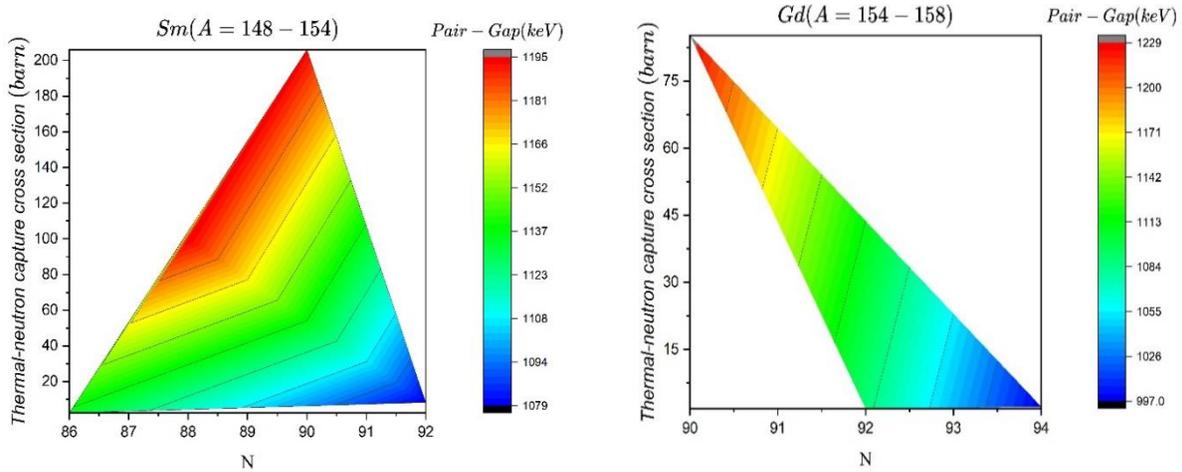

Figure 5. Contour plots in the ($\sigma_{th}$, neutron number) plane, in terms of the pairing gap in rare earth nuclei around A=150, including even-even Sm and Gd isotopic chains.

According to the conclusions of Refs.[39,40], the results for the cross-section of neutron capture indicate SPT in rare nuclei around A=150, and similarly, we investigated two isotopic chains in this region and, according to Fig. 5, our results about the pairing gap and $\sigma_{th}$ show a sensible variation around N=90. According to Fig. 5, this large peak in N=90 about the pairing gap in ($\sigma_{th}$, neutron number) plane can be related to changing phase to another one. This phenomenon can be a new application of the pairing gap to show the critical point symmetry. Therefore, we can conclude that N=90 is a critical point to this SPT in rare nuclei around A=150.

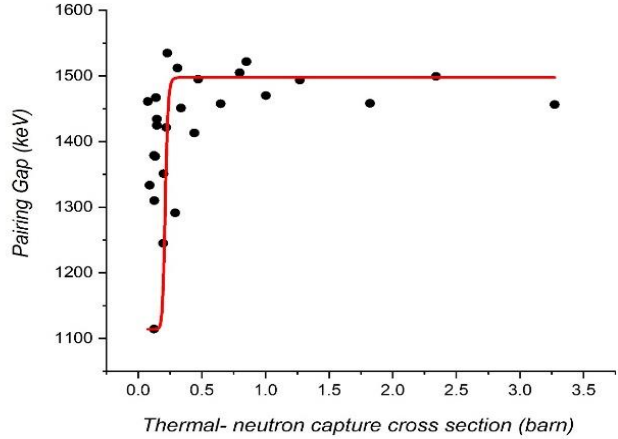

Figure 6. Experimental pairing gap values versus $\sigma_{th}$, for even-even isotopic chains in the near shell closure Z=50 (Ru, Pd, Cd, Sn)

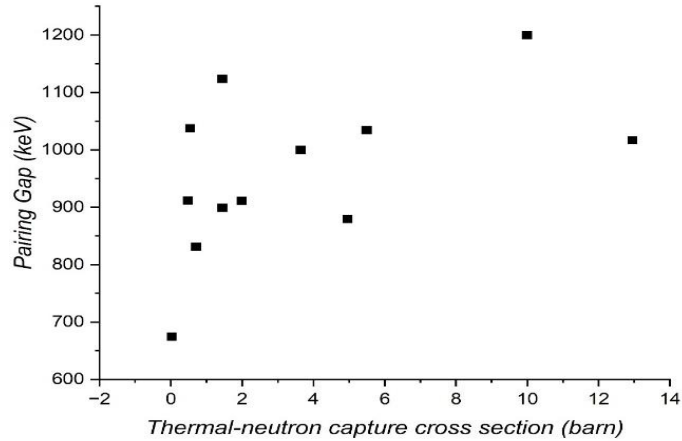

Figure 7. Experimental pairing gap values versus $\sigma_{th}$, for even-even isotopic chains in the near shell closure Z=82 (Os, Pt, Hg, Pb)

We can use a similar method that has been used in[39,40] for the correlation between $S_{2n}$ and $\sigma_{th}$, and knowing the pairing gap and $\sigma_{th}$ correlation in the near shell closure Z=50 and a fit function that describes it, given by the following:

$$\Delta_{\text{fit}} = A_2 + \frac{A_2 - A_1}{1 + \left(\frac{\sigma_{th}}{\sigma_{0th}}\right)^P}, \tag{10}$$

Where $A_1$ is 1114.12203±16.41994 (keV), $A_2$ is 1497.69199 ±5.59091 (keV), and P is 19.21272 ±4531.51.

According to the pairing gap's energy level and the cross-section ranges in Fig. 6 and Fig. 7, and compared to each other, similar global investigation about Fig. 3 and Fig. 4, we also can see the inverse relation between the pairing gap and $\sigma_{th}$ in Fig. 6 and Fig. 7. Eq.(10) also confirms this correlation. By increasing the pairing gap in Fig. 6, the number of isotopes that their $\sigma_{th}$ are in the range of 0-0.5 (barn) will be increased, but, after reaching the specific pairing gap's limit, the number of isotopes with upper than 0.5 (barn) will be decreasing. Similarly, this description is also valid for Fig. 7 but with a lower pairing gap energy level and a higher range of neutron capture cross-sections.

## 3. Conclusion

The quantum phase transition phenomena in different nuclei have been investigated by using the significant changes in the structure of different isotopic chains in which the neutron number has the role of control parameter. The two-neutron separation energies and the cross-section of thermal neutrons are known as the most sensitive observables to the nuclear structure, and their variations have been used to identify the QPT in different isotopic chains and also the nuclei, which are known as the candidates for critical point symmetries. This idea was the starting point of this investigation, and we tried to show that similar changes are happening in the values of the pairing gap in such isotopic chains of nuclei, which are located near the Z=50 and Z=82 proton closed shells and the QPT have been assigned for them. Our results show a closed correlation between the pairing gap and the two neutron separation energies for any nucleus in which the $S_{2n}$ values show different performance in comparison with other nuclei in the considered isotopic chains, the variation of the pairing gap has similar behavior. On the other hand, for the pairing gap and two-neutron separation energies, the same general correlation wasn't observed, and we explained such a relation for each of the considered isotopic chains individually. In the following, the correlation between the pairing gap and thermal neutron capture cross-section was investigated, and we concluded the inverse relation between these quantities. Also, identification of the critical point by the pairing gap in ($\sigma_{th}$, neutron number) plane as a new application of the pairing gap was proposed. These results suggest the pairing gap as a new observable to describe the QPT in different isotopic chains.

**Acknowledgment**

This work is supported by the Research Council of the University of Tabriz.

**Author contributions**

H. Emami and H. Sabri performed the initial calculations, analyzed and interpreted the results, and wrote the main manuscript text. All authors commented on and reviewed the manuscript.

**Competing interests**

The authors declare no competing interests.

**Data Availability Statement**

The datasets used and analyzed during the current study are available from the corresponding author at the reasonable request.